\documentstyle{article}

\def\beq{\begin{equation}}
\def\eeq{\end{equation}}

\begin{document}
\pagestyle{empty}
$\ $
\vskip 2 truecm

\centerline{\bf MASSLESS MINIMALLY COUPLED FIELDS IN DE SITTER SPACE:}
\centerline{\bf O(4)-SYMMETRIC STATES VERSUS DE SITTER INVARIANT VACUUM}

\vskip 1 truecm
\centerline{Klaus Kirsten$^{*,}$
\footnote{Present address: Universit\`a degli Studi di Trento,
Dipartimento di Fisica, 38050 Povo (Trento), Italy}
and Jaume Garriga$^{**}$}
\vskip .5 truecm
\centerline{\em $^{*}$Department of Physics, Universit\"{a}t Kaiserslautern,}
\centerline{\em  Erwin Schr\"{o}dinger
Stra$\beta$e, Postfach 3049, D-6750 Kaiserslautern, Germany}
\vskip .5 truecm
\centerline{\em $^{**}$Tufts Institute of Cosmology,}
\centerline{\em Department of Physics and Astronomy,
              Tufts University, Medford, MA 02155}

\vskip 2.5 truecm
\begin{abstract}

The issue of de Sitter invariance for a massless minimally coupled scalar field
is revisited. Formally, it is possible to construct a de Sitter invariant
state for this case provided that the zero mode of the field is quantized
properly. Here we take the point of view that this state is physically
acceptable, in the sense that physical
observables can be computed and have a reasonable interpretation.
In particular, we use this vacuum to derive
a new result: that the squared difference between the field at two
points along a geodesic observer's space-time path grows linearly with
the observer's proper time for a quantum state that does not break de
Sitter invariance.
Also, we use the Hadamard
formalism to compute the renormalized expectation value of the energy
momentum tensor, both in the O(4) invariant states introduced by Allen and
Follaci, and in the de Sitter invariant vacuum. We find that the vacuum
energy density in the O(4) invariant case is larger than in the de Sitter
invariant case.
\end{abstract}

\clearpage
\pagestyle{plain}
\section{Introduction}
\label{introduction}

Quantum field theory in curved spacetimes has been extensively studied
during the past two decades or so (see e.g. ref.\cite{bida82} for a review)
with the purpose of understanding quantum effects in the presence of strong
gravitational fields. In particular, a lot of attention has been devoted to
de Sitter space, mainly because it has a high degree of symmetry
and the wave equation  can be exactly solved in this background.
A 4-dimensional de Sitter space can be conveniently defined as a hyperboloid
embedded in a 5-dimensional Minkowski space:
\beq
\xi^{\mu}(x)\xi_{\mu}(x)=H^{-2},\label{loid}
\eeq
where $\xi^{\mu}(x)$ denotes the position vector of the point $x$ in the
embedding space $(\mu=0,...,4)$. The manifest invariance of
equation (\ref{loid})
under 5-dimensional Lorentz transformations implies that de Sitter space
has a 10 parameter group of isometries, the de Sitter group O(4,1).

Scalar fields of mass $m$ with an arbitrary coupling $\xi$
to the Ricci curvature
scalar [see eq. (\ref{action}) below] can be easily quantized in de Sitter
space, and quantum states that respect
the O(4,1) invariance of the background can be constructed for
these fields\cite{cheta68}.
Physical quantities such as the two point function and the
renormalized expectation value of the energy momentum tensor in the de
Sitter invariant states were
computed exactly in early work \cite{buda78}.

Later, interest
in this subject was motivated by the inflationary cosmology scenario
\cite{blgu}
(since the geometry of spacetime during inflation is that of de Sitter space).
In this context, it was realized that the mean squared fluctuations of a
massless minimally coupled field (i.e. $m=\xi=0$)
grow linearly with time during inflation \cite{vifo82},
\beq
<\phi^2>\approx{H^3\over 4\pi^2}t.\label{growth}
\eeq
Note that this expression is not de Sitter invariant, essentially because the
quantum state that was used in its derivation breaks the O(4,1) invariance
explicitly.

The massless minimally coupled case is peculiar in that the de Sitter
invariant two point function becomes infrared divergent in the limit
$m\to0,\ \xi\to0$. This led some authors \cite{al85,alfo87,po91}
to the definition of various other
`vacua' with less symmetry than the full de Sitter group,
but with a two point function which is free from infrared divergences. In
particular, here we will consider the two parameter family of
O(4)-invariant Fock `vacua' introduced by Allen and Folacci \cite{alfo87}.
Note that the $\xi^0=const.$ spatial sections of (\ref{loid}) are
3-spheres. The O(4) vacua are not invariant under all de Sitter
transformations, but only under spatial rotations of these 3-spheres.

In this paper we would like to reconsider the possibility of constructing a
de Sitter invariant state for the massless minimally coupled field.
(This state is not a Fock vacuum state.)
That such state can be formally constructed was already implicit in refs.
\cite{ra85,jackiw}, where the quantization was studied in the functional
Schr\"{o}dinger picture.
Our emphasis here will be in the physical interpretation.
For $m=\xi=0$ the action [eq.(\ref{action}) below] has a zero
mode: it is invariant under constant shifts of the field
$$
\phi\to\phi+const.
$$
The two point function is ill defined because all values of the
spatially constant part of the field are equally probable in the de Sitter
invariant state (which is analogous to an eigenstate of momentum in
the quantum mechanics of a free particle). However, such ambiguity
does not prevent us from
computing the expectation value of physical observables. To illustrate this
point here we shall use this vacuum
to derive a more powerful result than the one
given in equation (\ref{growth}).
Namely that one may have a de Sitter invariant state $|0>$, and in this
state, any freely-falling observer who picks a basepoint $x$ in
spacetime will see $<0|(\phi (x)-\phi (y))^2|0>$ increasing with
proper time along their path.
We shall also compute the renormalized expectation value of
the energy momentum tensor, both in the one-parameter family of O(4)
invariant states and in the de Sitter invariant vacuum.
As we shall see, the vacuum energy
density in the de Sitter invariant case is lower than in the
O(4)-symmetric case.

The rest of the paper is organized as follows. In Section 2 we briefly
review the quantization of a scalar field in de Sitter space, with the
purpose of fixing the notation. In Section 3 we compute the energy momentum
tensor in the two parameter family of O(4) invariant vacua. Section 4
discusses the de Sitter invariant vacuum for the massless minimally
coupled field. In Section 5 we use this vacuum for the calculation of some
observables.
Finally, a discussion of the results is given in Section 6.
The quantization of the scalar field
in the functional Schr\"{o}dinger picture is summarized in the Appendix.

\section{Scalar field in de Sitter space}

In this section we summarize the quantum theory of a scalar field
of mass $m$ and arbitrary coupling to the scalar curvature in de Sitter
space, which was
developed in Refs.\cite{cheta68,scsp76,buda78,al85,ra85}.

The line element in de Sitter space reads
\beq
ds^2=g_{ab}dx^adx^b=H^{-2}\sin^{-2}\eta[-d\eta^2+d\Omega^2],\label{metric}
\eeq
where we are using the closed coordinate system $x^a=(\eta,\Omega)$,
$(a=0,...,3)$ that covers the whole hyperboloid (\ref{loid}). Here
$\eta\in(0,\pi)$ is the so-called conformal time, $\Omega$ is a set of angles
on the 3-sphere and $d\Omega^2$ denotes the line element on the unit
3-sphere.

The action for the scalar field is given by
\beq
S={-1\over 2}\int\sqrt{-g}[\partial_a\phi\partial^a\phi+(m^2+\xi R)\phi^2]
d^4x, \label{action}
\eeq
where $g$ is the determinant of the metric, $R=12H^2$ is the Ricci scalar and
$\xi$ is an arbitrary coupling. It is convenient to expand the field as
\beq
\phi=\sum_{LM}\chi_{LM}(\eta){Y}_{LM}(\Omega),\label{phi}
\eeq
where ${Y}_{LM}$ are the
usual spherical harmonics on the 3-sphere, normalized as
\beq
\int {Y}_{LM}(\Omega){
Y}^*_{L'M'}(\Omega)d\Omega=\delta_{LL'}\delta_{MM'}. \label{norm}
\eeq
They are eigenfunctions of the Laplacian on the 3-sphere
\beq
\Delta^{(3)}{Y}_{LM}=-J{Y}_{LM},\label{laplace}
\eeq
with $J=L(L+2)$, $L=0,...,\infty$. The index $M$,
$M=0,...,(L+1)^2$, labels the degeneracy for given $L$.

Introducing (\ref{phi}) in (\ref{action}) one finds
\beq
S={1\over 2}\sum_{LM}\int (H\sin\eta)^{-2}[(\dot\chi_{LM})^2
-\omega_L^2(\eta)\chi^2_{LM}]d\eta,\label{action2}
\eeq
where
$$
\omega_L^2(\eta)\equiv J+{m^2+\xi R\over (H\sin\eta)^2},
$$
and the overdot indicates derivative with respect to $\eta$.
In going from (\ref{action}) to (\ref{action2}) the term $\partial_i{
Y}_{LM}\partial_j{Y}_{LM}$ has been integrated by parts and the
relations (\ref{norm}) and (\ref{laplace}) have been used.
Equation (\ref{action2}) can be seen as the action for a collection of
harmonic oscillators with time dependent frequencies. The classical
equations of motion for the modes $\chi_{LM}(\eta)$ read,
\beq
\ddot\chi_{LM}-2\cot\eta\dot\chi_{LM}+\omega_L^2(\eta)\chi_{LM}=0.
\label{eom}
\eeq

To quantize the theory, the field variables $\chi_{LM}$ and their
canonically conjugate momenta
\beq
\pi_{LM}\equiv{\partial L\over\partial\dot\chi_{LM}}=(H\sin\eta)^{-2}
\dot\chi_{LM},\label{pi}
\eeq
are promoted to operators satisfying the canonical commutation relations
\beq
[\hat\chi_{LM}, \hat\pi_{L'M'}]=i\delta_{LL'}\delta_{MM'}.\label{commutation}
\eeq
In the Heisenberg picture, these are time dependent operators, and it is
customary to expand them in terms of (time independent) creation and
anihilation operators $a_{LM}$ and $a^{\dagger}_{LM}$,
\beq
\hat\chi_{LM}=U_{LM}a_{LM}+U^*_{LM}a^{\dagger}_{LM}
\label{customary}
\eeq
$$
\hat\pi_{LM}=(H\sin\eta)^{-2}[\dot U_{LM}a_{LM}+\dot
U^*_{LM}a^{\dagger}_{LM}].
$$
Here $U_{LM}(\eta)$ are solutions of the field equation (\ref{eom}) (with
$\chi_{LM} \leftrightarrow U_{LM}$) normalized according to the Wronskian
condition
\beq
U_{LM}\dot U^*_{LM}-U^*_{LM}\dot U_{LM}=i(H \sin\eta)^2.\label{wronskian}
\eeq
The commutation relations (\ref{commutation})
follow from (\ref{wronskian}) and  the usual
commutation relations for the creation and anihilation operators
$$
[a_{LM},a^{\dagger}_{L'M'}]=\delta_{LL'}\delta_{MM'},\quad
[a_{LM},a_{L'M'}]=[a^{\dagger}_{LM},a^{\dagger}_{L'M'}]=0.
$$

A ``vacuum'' state $|0>$ can be defined by
\beq
a_{LM}|0>=0,\quad \forall L,M, \label{vacuum}
\eeq
and the complete Hilbert space of states can be generated by repeated
operation on
$|0>$ of the creation operators $a^{\dagger}_{LM}$. As it is usual in
curved space (see e.g. \cite{bida82}), the definition of this vacuum is
somewhat arbitrary, since it depends on what particular choice we make for
the set of modes $\{U_{LM}\}$. However, de Sitter space is a maximally
symmetric space, invariant under a 10 parameter group of isometries [the de
Sitter group O(4,1)], and it is natural to choose a vacuum state which also has
the same symmetry. Actually, there exists a one-parameter family of de Sitter
invariant quantum states. Among them, we shall concentrate on the so-called
Euclidean vacuum
as the only one whose two point function has Hadamard form and so the
ultraviolet behavior is the same as for field theory in flat
spacetime.
The mode functions corresponding to the Euclidean vacuum
are given by \cite{cheta68},
\beq
U_{LM}=A_L(\sin\eta)^{3/2}[P^{\lambda}_{\nu}(-\cos\eta)-{2i\over\pi}
Q^{\lambda}_{\nu}(-\cos\eta)],\label{modes}
\eeq
where $P^{\lambda}_{\nu}$ and $Q^{\lambda}_{\nu}$ are Legendre functions
on the cut,
and
\beq
\lambda=\left[{9\over4}-{{m^2+\xi R}\over H^2}\right]^{1/2},
\quad\nu=L+{1\over 2}.
\label{lambdasandnus}
\eeq
The normalization constants are given by
\beq
A_L={\sqrt\pi\over2}He^{i\lambda\pi/2}\left[{\Gamma(L-\lambda+3/2)\over
\Gamma(L+\lambda+3/2)}\right]^{1/2}.
\label{coefficients}
\eeq

The de Sitter invariance of this state is manifest in the symmetric two point
function
$$
G^{(1)}(x,x')=<0|\phi(x)\phi(x')+\phi(x')\phi(x)|0>=
$$
\beq
\sum_{LM}[U_{LM}(\eta)U^*_{LM}(\eta'){Y}_{LM}(\Omega){Y}^*_{LM}(\Omega')+
U_{LM}(\eta')U^*_{LM}(\eta){Y}_{LM}(\Omega'){Y}^*_{LM}(\Omega)]
,\label{sumovermodes}
\eeq
which can be evaluated to yield \cite{cheta68}
\beq
G^{(1)}(Z)={2H^2\over
(4\pi)^2}\Gamma\left({3\over2}-\lambda\right)\Gamma\left({3\over2}+\lambda
\right)
F\left({3\over2}-\lambda,{3\over2}+\lambda, 2;{1+Z\over
2}\right).\label{hypergeometric}
\eeq
Here $F$ is the hypergeometric function, and $Z$ is given by
\cite{alfo87}
\beq
Z(x,x')=H^2\xi^{\mu}(x)\xi_{\mu}(x')={\cos\gamma-cos\eta\cos\eta'\over
\sin\eta\sin\eta'},\label{zeta}
\eeq
where $\gamma$ is the angle between $\Omega$ and $\Omega'$. Note that the
two point function depends only on $Z$, which is a Lorentz invariant
quantity in the embedding space, and therefore $G^{(1)}$ is de Sitter
invariant.

The quantity $Z(x,x')$ can also be expressed as \cite{alfo87}
\beq
Z=\cos\sqrt{R\sigma\over 6},\label{zsigma}
\eeq
where $\sigma(x,x')$ is defined as one half of the square of the geodesic
distance between $x$ and $x'$. If $x$ and $x'$ are time-like separated,
then $\sigma<0$ and $Z>1$. On the other hand, if they are space-like
separated, then $Z<1$. (However a geodesic joining the two points
exists only if $-1 \le Z$, hence $\sigma$ is undefined for $Z<-1$.)

\section{The massless minimally coupled case: O(4) invariant vacuum}

It should be noted that the two-point function (\ref{hypergeometric}) is ill
defined in the massless minimally coupled case $(m=\xi=0)$, since one of
the gamma functions has a pole at $\lambda=3/2$. This divergence has led
some authors \cite{al85,alfo87,po91}  to the definition of other vacua with
less symmetry than the full de Sitter group, but with a well defined two point
function.

In the closed coordinate system that we are using, one such natural vacuum
is the O(4) invariant vacuum \cite{alfo87}, which is symmetric under
rotations of the $\eta=const.$ spatial sections (which are 3-spheres).
The set of modes that defines the O(4)
invariant quantum state is given
by (\ref{modes}) for $L>0$ but, in order to avoid the infrared
divergence, the $L=0$ mode solution is chosen as $\cite{alfo87}$
\beq
U_0=H[A(\eta-{1\over 2}\sin2\eta-{\pi\over 2})+B],
\label{altarriba}
\eeq
with
$$
A=-i\alpha,\quad\alpha\in(0,\infty),
$$
$$
B={1\over \alpha}({1\over 4}+i\beta),\quad\beta\in(-\infty,\infty).
$$
The two complex parameters $A$ and $B$ have been reduced to two real
parameters $\alpha$ and $\beta$ because an overall phase is irrelevant and
because (\ref{wronskian}) must be satisfied. In addition, requiring
time reversal invariance fixes $\beta=0$ \cite{alfo87}, which leaves us
with just one parameter, $\alpha$. In what follows we take $\beta=0$.

The two point function in this state is
\beq
G^{(1)}_{\alpha}(x,x')=\hat G(x,x')+{1\over
2\pi^2}[U_0(\eta)U_0^*(\eta')+U_0(\eta')U_0^*(\eta)],
\label{g1ab}
\eeq
where $\hat G$ is defined as a sum over modes
similar to (\ref{sumovermodes}) but without the $L=0$ term. This sum is given
(up to some irrelevant constant) in closed form by \cite{alfo87}
\beq
\hat G(x,x')={R\over 48\pi^2}\left[{1\over
1-Z}-\log(1-Z)-\log(4\sin\eta\sin\eta')-\sin^2\eta-\sin^2\eta'\right],
\label{hatg}
\eeq
with $Z$ defined in (\ref{zeta}).

We will be interested in constructing the energy momentum tensor using the
Hadamard formalism \cite{befo86,brot86}.
For
this we need to study the two point function in the coincidence limit, that is,
we have to bring $G^{(1)}$ into the Hadamard form \cite{fr75,ha23}
\beq
G^{(1)}(x,x')={1\over 4\pi^2}\left[
{\Delta^{1/2}(x,x')\over\sigma}+V(x,x')\log\sigma+W(x,x')\right],
\label{hadamard}
\eeq
where $\sigma(x,x')$ was defined in eq. (\ref{zsigma})
and $\Delta(x,x')$ is the Van Vleck-Morette determinant. In de Sitter
space it is given by \cite{fr75}
\beq
\Delta(\sigma)=\left({R\sigma\over
6}\right)^{3/2}\left[\sin\sqrt{R\sigma\over6}\right]^{-3}. \label{vvleck}
\eeq
In eq.(\ref{hadamard}), $V(x,x')$ and $W(x,x')$ are symmetric functions of $x$
and $x'$ which are smooth in the coincidence limit.

Using (\ref{zsigma}) and (\ref{vvleck}), one can compare expressions
(\ref{hadamard}) and (\ref{g1ab}-\ref{hatg}) to find
\beq
V(x,x')=-{R\over 12},\label{V}
\eeq
$$
W(x,x')=F(\sigma)-{R\over
12}[\log(4\sin\eta\sin\eta')+\sin^2\eta+\sin^2\eta']+
$$
\beq
2[U_0(\eta)U^*_0(\eta')+U^*_0(\eta)U_0(\eta')].
\label{W}
\eeq
Here
\beq
F(\sigma)\equiv {R\over 12}\left[{1\over 1-\cos X}
-2{1\over(X\sin^3X)^{1/2}}-\log\left({R\over 6X^2}[1-\cos X]\right)\right],
\label{F}
\eeq
with
$$
X=\sqrt{R\sigma\over 6}.
$$
One can check that $W$ is well behaved at $\sigma=0$ (as expected from the
general theory) by
expanding each term in (\ref{F}) in powers of $\sigma$. We find
that the negative powers of $\sigma$ cancel out and we have
\beq
F(\sigma)=-{R\over12}\left[\log\left({R\over 12}\right)+{1\over3}-
{R\sigma\over480}+ O(\sigma^2)\right].\label{powers}
\eeq
As usual, the singular part in (\ref{hadamard}) is purely geometrical, and
all the dependence
of $G^{(1)}$ on
the quantum state is contained in the function $W(x,x')$.

The two point function is now in a form ready for the computation of
the renormalized expectation value of the energy momentum tensor.
Using the Hadamard formalism, this is given by \cite{befo86,brot86}
\beq
8\pi^2<T_{ab}>_{ren}={\cal\tau}_{ab}[W]-{\cal\tau}_{ab}[V]\log\mu^2+
2v_1g_{ab}-{m^4\over 16}g_{ab},
\label{formula}
\eeq
where
$$
{\cal\tau}_{ab}[f]\equiv\lim_{x\to x'}{\cal D}_{ab'}(x,x')[f(x,x')].
$$
Here ${\cal D}$ is the differential operator associated with the point
splitted expression of the formal energy-momentum operator. In the massless
minimally coupled case
$$
{\cal D}_{ab'}\equiv \nabla_a\nabla_{b'}-{1\over
2}g_{ab'}g_{dd'}\nabla^d\nabla^{d'},
$$
where $g_a^{b'}$ is the bivector of parallel transport \cite{dewitt63}. In
eq.(\ref{formula}), $\mu^2$ is a renormalization scale (arbitrary, in
principle), and $v_1$ is the `trace anomaly' scalar,
which in de Sitter space is
equal to \cite{brot86}
$$
v_1={29 R^2\over 8640}.
$$

{}From (\ref{V}) it is clear that in our case
$$
{\cal\tau}_{ab}[V]=0,
$$
and the dependence on the renormalization scale disappears. This is
fortunate, since in the massles case there is no natural mass parameter in the
problem. Also, the last term in (\ref{formula}) vanishes for $m=0$.

All that we need to evaluate is ${\cal\tau}_{ab}[W]$, with $W$ given by
(\ref{W}). The term ${\cal\tau}_{ab}[F(\sigma)]$ can be easily computed by
noticing that
\beq
\lim_{x\to x'}\nabla_a\nabla_{b'}F(\sigma)=
\lim_{x\to x'}[F''(\sigma)\sigma_{,a}\sigma_{,b'}+F'(\sigma)\sigma_{;ab'}]=
-F'(\sigma)|_{\sigma=0}g_{ab}, \label{ajut}
\eeq
where a prime indicates derivative with respect to $\sigma$ and we have
used (see e.g.\cite{befo86})
$$
\lim_{x\to x'}\sigma_{,a}=0,
$$
$$
\lim_{x\to x'}\sigma_{ab'}=-g_{ab}.
$$
The value of $F'(\sigma)$ at $\sigma=0$ can be read off from (\ref{powers}),
and using (\ref{ajut}) we have
$$
{\cal\tau}_{ab}[F]={R^2\over 5760}g_{ab}.
$$
Also, it is clear that
$$
{\cal\tau}_{ab}[\log(2\sin\eta)+\log(2\sin\eta')+\sin^2\eta+\sin^2\eta']=0,
$$
and one can check that
$$
{\cal\tau}_{ab}[U_0(\eta)U^*_0(\eta')+U_0^*(\eta)U_0(\eta')]=
{1\over36}R^2{\alpha}^2(1-2\delta_{a\eta})\sin^6\eta\ g_{ab}.
$$

Substituting the previous expressions in (\ref{formula}), we have
\beq
<\alpha|T_{ab}|\alpha>_{ren}={119R^2 \over 138240\pi^2}g_{ab}+
{R^2\over 144\pi^2}
{\alpha}^2\sin^6\eta\ g_{ab}(1-2\delta_{a\eta}).\label{tmunu}
\eeq
Therefore, the energy momentum tensor is not de Sitter invariant, but only
O(4) invariant, because of the explicit time dependence. Notice also
that the term which is not de Sitter invariant decays
with the expansion of the Universe as $a^{-6}$,
where $a$ is the scale factor (compare with radiation,
which behaves as $a^{-4}$ or with the vacuum energy itself which
behaves as $a^{0}$)
and therefore it is unlikely to have
any cosmological consequences. In the limit $\eta\to 0$ or
$\eta\to\pi$, which
corresponds to cosmological time going to $+\infty$ or $-\infty$,
eq. (\ref{tmunu}) reduces to
the result (3.6) in Ref.\cite{alfo87}
as corrected in Ref.\cite{referee}.

\section{De Sitter invariant vacuum for the massless minimally coupled case}

Sometimes it is said \cite{al85,po91}
that the infrared divergence in $G^{(1)}$ indicates that
de Sitter invariance is broken in the massless minimally coupled case.
However,
it is still possible to define a de Sitter invariant vacuum for this case,
and here we will take the point of view that this state is
physically acceptable in the sense that physical quantities can be computed
and have a reasonable interpretation.
However, as we shall see, the space of states can not be
simply represented as a Fock
space built by applying creation operators to this vacuum state.
The quantization of $\phi$ in the case $m=\xi=0$ is peculiar because the field
contains a zero mode: the action is invariant under the transformation
\beq
\phi\to\phi+const.\label{transformation}
\eeq
It is well known that an expansion in terms of
creation and anihilation operators, such as (\ref{customary}), is not
adequate for the variables associated to the zero modes
\cite{dewitt83,fopa89,gavi91b,rajaraman}.

The situation is
analogous to that of a quantum mechanical harmonic oscillator: the
expansion of the position and momentum operators, $x$ and $p$,
in terms of creation and
anihilation operators
breaks down in  the limit when the frequency
of the oscillator, $\omega$, goes to zero (the free particle case).
In the Heisenberg picture we have
$$
x(t)=(2M\omega)^{-1/2}(ae^{-i\omega t}+ a^{\dagger}e^{+i\omega t})
$$
$$
p(t)=-i(M\omega/2)^{1/2}(ae^{-i\omega t}- a^{\dagger}e^{+i\omega t}),
$$
where $M$ is the mass of the particle.
Of course, these expressions are not valid in the limit $\omega\to 0$.
The physical reason is that for a free particle
the spectrum of the Hamiltonian becomes continuous and the number operator
loses its meaning.
Instead, we can consider the expansions
\beq
x(t)=x_0+p_0 t \label{canonge}
\eeq
$$
p(t)=p_0,
$$
where the new operators satisfy the commutation relation $[x_0,p_0]=i$. At
the classical level, $x_0$ and $p_0$ have the interpretation of the initial
position and momentum (and are therefore constants) so (\ref{canonge}) can
be seen as a Hamilton-Jacobi canonical transformation in which the new
variables are constants of motion. The first equation in (\ref{canonge}) is
obviously the general solution of the equations of motion if we think of
$x_0$ and $p_0$ as constants of integration. In this sense this equation is
analogous to (\ref{altarriba}).

The simplest example of a field theory with zero modes is the massless
scalar field in a flat compact space with finite volume $V$ and topology of
a torus $S^1\times S^1\times S^1$, discussed in Ref. \cite{fopa89}.
In that case, a complete set of solutions of the wave equation is given by
\beq
f_{\bf k}=(2V\omega)^{-1/2}\exp i({\bf kx}-\omega t),
\quad ({\bf k}\neq 0)\label{edw}
\eeq
$$
f_0=At+B.
$$
Here $\omega=|{\bf k}|$ and the momenta ${\bf k}$ have the usual discrete
spectrum due to finite volume. The Klein-Gordon normalization
requires $A^*B-B^*A=i/V$. While the modes $f_{\bf k}$ (${\bf k}\neq 0$)
are the classical solutions for a harmonic oscillator of frequency
$\omega$, the mode $f_0$ is the classical solution for a free particle.
Therefore, although it is formally possible to define creation and
anihilation operators associated with $f_0$, in a manner analogous to the
construction of the $O(4)$ invariant vacuum of the previous section, it is
more natural to define position and momentum operators analogous to $p_0$
and $x_0$ above. With this the field expansion reads \cite{fopa89}
$$
\phi={x_0+p_0 t\over\sqrt V}+\sum_{{\bf k}\neq 0}(a_{\bf k}f_{\bf k}+h.c.).
$$
It can be checked that the equal time commutation relation for $\phi$ and
its conjugate momentum are satisfied if $[x_0,p_0]=i$ and the
usual commutation relations for the creation and anihilation operators are
satisfied.

Note that in the limit of infinite volume the special treatment of the zero
mode becomes irrelevant, as it makes a contribution of zero measure in the
expansion of the field. An equivalent statement is that the set of modes
with ${\bf k}\neq 0$ becomes complete in the limit of infinite volume.
However, for finite volume the zero mode is important and makes a finite
contribution to the energy. Indeed, it is straightforward to see that
$$
E={p_0^2\over 2}+\sum_{\bf k}|{\bf k}|\left(a_{\bf k}^\dagger a_{\bf
k}+{1\over 2}\right).
$$
One can define the ground state for this system through the equations
$p_0|0>=0$, $a_{\bf k}|0>=0$. This ground state is not normalizable, in the
same way that the ground state of a quantum mechanical free particle is
not (for a detailed discussion on this issues, see Ref.\cite{dewitt83},
Section 9). The field operator is seen to be equivalent to a collection of
harmonic oscillators plus a free particle [whose position, in the
Heisenberg picture, would be given by the operator $x(t)=x_0+p_0 t$]. The
space of states is equivalent to the direct product of a Fock space
corresponding to the oscillators and an ordinary Hilbert space
corresponding to the free particle. Since the energy is an observable, in
addition to the usual Fock space operators, the momentum $p_0$ is also an
observable.

The above construction can be generalized to arbitrary curved backgrounds
\cite{dewitt83}. Of course, in general, there is the usual caveat that for
non-stationary backgrounds the energy is not conserved and the definition
of a ground state is ambiguous. This is nothing new, it is the same problem
that we encountered in Section 2 when discussing the massive field: the
definition of a ``vacuum" in non-stationary backgrounds is always a matter
of choice. Here, as in Section 2, we will be guided
by considerations of symmetry in making this choice.

In the case of a massless minimally coupled field in De Sitter space,
the zero mode associated to (\ref{transformation}) is in the
homogeneous sector $(L=0)$, and that is the reason why the coefficient
$A_0$ [see eq. (\ref{coefficients})] becomes infinite for $\lambda\to3/2$.
Instead of defining creation and anihilation operators for $L=0$
we replace the expansions (\ref{customary}) by
$\cite{alfo87}$
\beq
\chi_{0}={H\over\sqrt{2}}[Q+(\eta-{1\over2}\sin 2\eta-{\pi\over 2})P]
\label{jacobi}
\eeq
$$
\pi_0={\sqrt 2\over H}P.
$$
The coefficients of $Q$ and $P$ in the expansion of $\chi_0$ are solutions
of the field equation (\ref{eom}), and the expression for $\pi_0$ follows
from (\ref{pi}). Moreover,
the commutation relation between $\chi_0$ and $\pi_0$ implies
$$
[Q,P]=i,
$$
so, again, (\ref{jacobi}) can be seen as a Hamilton-Jacobi transformation
in which the new canonical variables are constants of motion.

We define a vacuum state by
\beq
P|0>=0,\label{p0}
\eeq
$$
a_{LM}|0>=0,\quad L>0,
$$
where $a_{LM}$ were defined in Section 2.

The ambiguity in the choice of a vacuum corresponds to the freedom in the
choice of the mode functions $U_{LM}$ for $L\neq 0$ [which we take to be the
same as for the O(4) vacuum], plus the freedom in choosing the mode
solutions which appear as coefficients of $Q$ an $P$ in Eq.(\ref{jacobi}).
In principle we could have chosen any two homogeneous solutions of the
wave equation, say $f_1(\eta)$ and $f_2(\eta)$,
$$
\chi_0=f_1\tilde Q+f_2\tilde P
$$
$$
\pi_0=(H\sin\eta)^{-2}(\dot f_1\tilde Q+\dot f_2\tilde P),
$$
subject to the Wronskian condition $\dot f_2f_1-\dot f_1f_2=H^2\sin^2\eta.$
With the choice (\ref{jacobi}) the equation $P|0>=0$ implies that
the vacuum wave functional $\Psi$ does not depend on
$\chi_0$
\beq
P\Psi={H\over\sqrt{2}}\left(-i{\partial\over\partial\chi_0}\right)\Psi=0.
\label{independent}
\eeq
If we are interested in a de Sitter invariant vacuum, this turns out to be
the right choice.

In appendix A we review the quantization of the scalar field in
the Schr\"{o}dinger picture.
We show that in the limit $m\to0$ and $\xi\to0$, the de
Sitter invariant wave functional becomes independent of $\chi_0$, and
therefore it satisfies $P|0>=0$ (the other equations in (\ref{p0})
are also satisfied by construction). Note that the solution of
(\ref{independent}) is not normalizable, and that is the reason why
$G^{(1)}$ is ill defined in the de Sitter invariant state. This
should not be taken as an indication that the state is pathological:
it simply  means that all values of $\chi_0$ are equaly probable.
The same problem would arise in the quantum mechanics of a free particle if
we tried to compute $<p|x^2|p>$, where
$|p>$ is an eigenstate of momentum.

Apart from considerations about De Sitter invariance
(the group of isometries of the
background spacetime), there is another (aesthetic) reason for choosing a
state with $P|0>=0$, based on the symmetry of the Lagrangian
under $\phi\to \phi+const.$
The corresponding Noether
current is $j_{\mu}=\partial_{\mu}\phi$.
The generator of the symmetry is
the ``charge''
$$
\hat Q=\int_{\eta=const} d\Sigma_{\mu}j^{\mu},\label{charge}
$$
so the vacuum will be invariant under this symmetry if it is anihilated
by the charge, $\hat
Q|0>=0$. Introducing $j_{\mu}=\partial_{\mu}\phi$ in (\ref{charge}) we find
$\hat Q=2\pi H^{-1}P$, so the condition becomes $P|0>=0.$ Note that even
though the current is linear in $\phi$, the charge operator is
non-vanishing and well defined precisely because the space has compact
spatial sections.

As mentioned before, the vacuum state defined in this way is not simply a
Fock-space vacuum (in fact, this would be in contradiction to the work
of Allen \cite{al85})
but the direct product of a Fock space and an ordinary Hilbert
space corresponding to the $\chi_0$ variable.
As we shall see, in order that the energy density
$T_{00}$ is a physical observable, in addition
to the usual Fock space observables the operator $P$ is
also a physical observable.

A basis for the space of states is the direct product of the basis for the
Fock space
times the basis for the Hilbert space
of a particle in one dimension. The structure of the Fock space
corresponding to the modes $L>0$ is identical to the one corresponding to
the 0(4) invariant vacuum, and we shall not discuss it further.
The Hilbert space for one particle in one dimension is isomorphic to the
usual space of square integrable complex functions of a real variable,
and a convenient basis is formed by the eigenstates of the momentum operator
$P$ with eigenvalue $p$
$$
|p>\equiv e^{ipQ}|0>.
$$
Since they form a continuous basis, these states are not normalized
in the discrete sense, but they have the continuous normalization
$<p|p'>=\delta(p-p')$. In the `q' representation they are the ordinary
plane waves
\beq
<q|p>=(2\pi)^{-1/2}e^{ipq}\label{pws}
\eeq
where $|q>$ are eigenstates of $Q$ with eigenvalue $q$,
normalized as $<q|q'>=\delta(q-q')$.

In this representation $Q$ acts as a multiplicative operator and $P$ as a
derivative operator
$$
<q|Q|\psi>=q<q|\psi>,
$$
$$
<q|P|\psi>=-i{\partial\over \partial q}<q|\psi>.
$$

To make a connection with the previous section, one can see that
the $L=0$ sector of the O(4) invariant vacuum $|\alpha>$ discussed previously
corresponds to the normalized gaussian wave packet \cite{alfo87}
\beq
<q|\alpha>\equiv \psi_{\alpha}(q)=
{\sqrt{2\alpha}\over\pi^{1/4}}e^{-2\alpha^2q^2}.\label{packet}
\eeq
Indeed, the operator $a_0$ of the O(4) vacuum can be expressed
[using (\ref{jacobi})] as $a_0=i\sqrt{2}[B^*P-A^*Q]$, which clearly
anihilates (\ref{packet}).
Also, the ``multi-particle'' homogeneous ($L=0$) excitations above
$|\alpha>$ are obtained by repeated operation of $a^{\dagger}_0$ on
(\ref{packet}), which gives,
\beq
\psi^n_{\alpha}\equiv <q|{(a^{\dagger}_0)^n\over\sqrt{n!}}|\alpha>=
\left({2\alpha\over\pi^{1/2}2^n n!}\right)^{1/2}H_n(2\alpha q)
e^{-2\alpha^2q^2},\label{hermits}
\eeq
where $H_n$ are the Hermite polynomials.

Throughout this section we have worked in the Heisenberg picture, and
therefore the states (\ref{hermits}) are time independent. To obtain the
corresponding wave functions in the Schr\"{o}dinger picture
one can solve the Schr\"{o}dinger
equation with initial conditions (\ref{hermits}). As we show in
Appendix A, this Schr\"{o}dinger equation is just the one for a free particle,
so the evolution of (\ref{packet}) is just that of a minimal wave packet which
spreads in time.

\section{Dispersion of the field and energy momentum tensor}

In order to gain intuition on the structure of the
de Sitter invariant vacuum defined in (\ref{p0}), let
us consider the `dispersion' of the field, defined by
\beq
D^2(x,y)\equiv<0|(\phi(x)-\phi(y))^2|0>,\label{distortion}
\eeq
which will give us an idea on how the value of the field fluctuates over space
and time. Since $D^2$ contains terms of the form $<0|\phi(x)\phi(x)|0>$,
we will encounter the usual ultraviolet divergences
associated with the product of operators in the coincidence limit. A
convenient way of getting around such divergences is to smear the field
operator over a region of size $s$ (see e.g. \cite{collins,vi83})
$$
\phi_s(x)\equiv{1\over Vol(s)}\int_{d(x,x')<s/2}\phi(x')d^3x',
$$
where $d(x,y)$ is the geodesic distance between $x$ and $x'$ and $Vol(s)$ is
the volume of the smearing region.
Here, and for the rest of this section,
$d^3x$ stands for the three dimensional invariant volume element.
The `diameter' $s$ of the smearing region
should be less than $2\pi H^{-1}$, since there are no space-like geodesics
longer than that [see comments after eq. (\ref{zsigma})], and we
shall take $s\sim H^{-1}$. Also, in order to smear the field
operator it is necessary to make a particular choice of space-like
hypersurface at the point $x$. In what follows we shall always
consider situations in which geodesic observers are involved, so
the smearing regions can be defined on the space-like sections
orthogonal to these geodesics. For instance, if $x$ and $y$ are time-like
separated, we can consider the geodesic
curve that links $x$ with $y$, and take
space-like surfaces at $x$ and $y$ generated by the space-like geodesics
orthogonal to this curve. Later, we shall also consider the field measured
by two observers moving along two different geodesics. Each observer can
smear the field on the space-like surface orthogonal to his or her
geodesic. In any case, in the limit of large separation between $x$ and
$y$, the leading term in the dispersion will not depend on the details of
how we smear the field.

Now we can
consider the dispersion of the smeared field,
\beq
D_s^2(x,y)=<0|(\phi_s(x)-\phi_s(y))^2|0>.\label{distortions}
\eeq
Notice that this expression has no infrared divergences either,
since the operator
$Q$ (which causes trouble in the two point function because its
expectation value is ill defined in the de Sitter invariant vacuum) cancels
out when we consider the difference $\phi(x)-\phi(y)$.

As an intermediate step to compute (\ref{distortions}) we `point-split' and
symmetrize the expression (\ref{distortion})
$$
D_{\epsilon}^2(x',x'';y',y'')\equiv{1\over2}<0|\{[\phi(x')-\phi(y')],
[\phi(x'')-\phi(y'')]\}|0>.
$$
Here $x'$ and $x''$ are points within the smearing region surrounding $x$,
separated by a geodesic distance $\epsilon$ $(\epsilon<s)$
(similarly for $y'$ and $y''$), and the brackets $\{,\}$ denote the
anticommutator. Since $P|0>=0$, this expression reduces to
$$
D^2_{\epsilon}={1\over2}[\hat G(x',x'')+\hat G(y',y'')-\hat G(x',y'')
-\hat G(y',x'')],
$$
with $\hat G$ defined in (\ref{g1ab}-\ref{hatg}). It is convenient to rewrite
it as
\beq
D_{\epsilon}^2={H^2\over 8\pi^2}[g(x',x'')+g(y',y'')-g(x',y'')-g(y',x'')],
\label{distortione}
\eeq
where
$$
g(u,v)\equiv{1\over 1-Z(u,v)}-\log|1-Z(u,v)|.
$$
Notice that (\ref{distortione}) is a fully de Sitter invariant expression
(as it should, since we are dealing with a de Sitter invariant state). All
the terms in (\ref{hatg}) that depend explicitly on the conformal time
$\eta$ cancel out in the expression for $D_{\epsilon}$.

Now we can easily estimate the smeared dispersion $D^2_s(x,y)$ for the case
when the separation between $x$ and $y$ is much larger than
$H^{-1}$ (that is $|Z|>>1$). This can be done by smearing
the expression
(\ref{distortione}) term by term. We note that when the two points
$x'$ and $x''$ lie within the same smearing region, the integrals
$$
{1\over [Vol(s)]^2}\int_{d(x,x')<s/2}d^3x'\int_{d(x,x'')<s/2}d^3x''
g(x',x'')
$$
give contributions of order 1, while if one of the points lies in the
neighborhood of $x$ and the other lies in the neighborhood of $y$, we have
$$
{1\over [Vol(s)]^2}\int d^3x'd^3y''g(x',y'')\approx g(x,y)\approx-\log
|Z(x,y)|,
$$
where we have used $|Z|>>1$. As a result,
\beq
D_s^2(x,y)\approx{H^2\over 4\pi^2}\log |Z(x,y)|,\quad(|Z|>>1).\label{dsz}
\eeq
If $x$ and $y$ are timelike separated, we can use (\ref{zsigma}) to write
\beq
<0|(\phi_s(x)-\phi_s(y))^2|0>\approx{H^3\over 4\pi^2}\tau,\quad (\tau>>H^{-1})
\label{result}
\eeq
where $\tau$ is the proper time measured by a geodesic observer travelling
from $x$ to $y$.

Equation (\ref{result}) embodies a familiar property of massless minimally
coupled fields in de Sitter space, namely, that the mean squared
fluctuations in the field grow linearly with time \cite{vifo82,vi83}
[see eq.(\ref{growth})]. Here
we have been able to derive this result in an invariant way,
without the need of
using a quantum state that breaks de Sitter invariance and without the need
of introducing a cosmological time coordinate [$\tau$ in eq.(\ref{result}) is
just the geodesic distance].
As noted by Vilenkin \cite{vi83}, the linear growth in time of the mean
squared fluctuation can be interpreted in terms of a random walk of the
field $\phi$. The magnitude of $\phi$ smeared over the
interior of a Hubble-radius ($H^{-1}$) two-sphere
changes by $\pm (H/2\pi)$ per expansion time $H^{-1}$. Then, the average
displacement squared is $D^2_s\sim (H/2\pi)^2 N$, where $N\sim H\tau$ is
the number of steps. To support this interpretation, Vilenkin studied field
correlations between points which were at large space-like separations.
We can repeat his arguments using the de Sitter invariant formalism.

Since points separated by space-like
distances greater than $\pi H^{-1}$ cannot be
connected by geodesics (and we are interested in much larger separations),
the discussion will require more work than in the case of
time-like separations. Consider, to begin with, an arbitrary point $x$ in de
Sitter space and a time-like geodesic $C_x$ passing through it. We can
think of $C_x$ as the trajectory of an inertial observer.  Without loss
of generality (by using de Sitter transformations)
we can take $x$ to have coordinates
$(\eta=\pi/2,\Omega)$, and $C_x$ to be the curve $\Omega=const.$, while the
metric still takes the form (\ref{metric}).
Let $x'$ be a
second point on the spacelike hypersurface orthogonal to $C_x$ at $x$, such
that the geodesic distance between $x$ and $x'$ is much {\em smaller} than
$H^{-1}$; and let $C_{x'}$ be a geodesic through the point $x'$ which is
initially parallel to $C_x$. In our coordinate system,
$x'$ has coordinates
$(\eta=\pi/2, \Omega')$, $C_{x'}$ is the curve $\Omega'=const.$,
and the distance between $x$ and $x'$ is $\gamma
H^{-1}$, ($\gamma<<1$), where $\gamma$ is the angle between $\Omega$ and
$\Omega'$.  Parametrizing both geodesics by the proper time
$\tau$ and taking $\tau=0$ at $\eta=\pi/2$, we can find what is the
separation between points in $C_x$ and $C_x'$ at any given $\tau$.

{}From (\ref{zeta}) we have,
$$Z(\tau,\gamma)=1+[\cos\gamma-1]\cosh^2H\tau,$$
where $Z(\tau,\gamma)$ means the invariant function $Z$ between the two points
on the geodesics $C_x$ and $C_x'$ at proper time $\tau$ and we have used
$(sin\eta)^{-1}=\cosh H\tau$. Two observers at $x$ and $x'$ which were
initially close and at rest relative to each other ($Z\approx 1$), are pulled
apart by the
expansion, so that eventually they reach large space-like separation
$Z<<-1$. The distance between both observers will be equal to $(\pi/2)H^{-1}$
at the time $\tau_*$ when $Z(\tau_*,\gamma)=0$ [see eq.(\ref{zsigma})], so
we can write
$$
Z(\tau,\gamma)=1-{\cosh^2H\tau\over \cosh^2H\tau_*}.
$$
For $\tau>>\tau_*$ we have [from(\ref{dsz})]
\beq
D_s^2\approx 2\cdot{H^3\over4\pi^2}(\tau-\tau_*).\label{dssp}
\eeq

In the language of Ref.\cite{vi83}, this result can be phrased as follows.
The field measured by each one of the two observers undergoes a random walk
of step $\Delta\phi_s=\pm(H/2\pi)$. As long as both observers lie within
the same Hubble volume their steps are correlated and the dispersion of
the field does not grow. Aproximately after time $\tau_*$, the
Hubble volumes around the two observers
stop overlapping, this means
the future light cones of the two observers fail to
overlap
and so the respective random walks of the field
become uncorrelated. Therefore, the dispersion is proportional to
$(\tau-\tau_*)$. The factor of $2$ in Eq. (\ref{dssp}) arises because we
have two independent random walks.

Finally we should say that although (\ref{result}) and (\ref{dssp}) have
been derived using the de Sitter invariant state, they would hold for any
O(4) invariant state (in the limit of large $\tau$). This is because the
contribution of $L=0$ to $D_s^2$ is
$$
{H^2\over 2\pi}(\eta-{1\over 2}\sin2\eta+(\eta\leftrightarrow
\eta')^2)<P^2>.
$$
This term remains bounded in time and eventually becomes subdominant with
respect to the vaccuum terms (\ref{result}) and (\ref{dssp}). Similarly,
because the modes $U_{LM}$ are bounded in time, any finite number of particles
in the modes $L>0$ will make a bounded contribution which will be irrelevant at
late times.

Another physical quantity that we can compute using $|0>$ is the
expectation value of the energy-momentum tensor. Since the
differential operator ${\cal D}_{ab'}$ acting on a constant is zero, the
operator $Q$ will not be present in the formal expression of $T_{ab'}$.
Also, since $P|0>=0$, it is clear that
$$
<0|{\cal D}_{ab'}\{\phi(x),\phi(x')\}|0>={\cal D}_{ab'}\hat G(x,x').
$$
The computation of $<0|T_{ab}|0>$ now reduces to the one presented in
Section 4, replacing $G^{(1)}_{A,B}$ by $\hat G$. Obviously, the result
is given by eq.(\ref{tmunu}) with $A=0$
\beq
<0|T_{ab}|0>_{ren}={119\over 138240\pi^2}R^2g_{ab},\label{caesar}
\eeq
which is de Sitter invariant as expected.

Since we chose a state with $P|0>=0$ there is no contribution from the $L=0$
sector to $<T_{ab}>$. The $L=0$ contribution to the energy
momentum tensor operator is
$$
\hat T^{(L=0)}_{ab}={R^2\over 144\pi^2}(1-2\delta_{a0})\sin^6\eta\ g_{ab}
{\hat P^2\over 2}.
$$
In a state with non-vanishing momentum, the expectation value of this
operator has to be added to the r.h.s. of (\ref{caesar}). In particular for
the O(4) invariant states $<P^2>_{\alpha}=2\alpha^2$ and we recover
(\ref{tmunu}). Clearly, for the energy $<T_{00}>$ to be an observable,
$P$ has to be observable.

It is interesting to compare
eq. (\ref{caesar}) with the general result for a massive and non-minimally
coupled field \cite{buda78},
$$
<T_{ab}>_{ren}={-g_{ab}\over64\pi^2}\{m^2[m^2+(\xi-{1\over6})R]
\left[\psi\left({3\over2}-\lambda\right)+\psi\left({3\over2}+\lambda
\right)+\log{R\over12m^2}\right]-
$$
$$
m^2\left(\xi-{1\over6}\right)R-{1\over18}m^2R-{1\over2}
\left(\xi-{1\over6}\right)^2R^2+{R^2\over2160}\ \}.
$$
Notice that the limit of this expression as $m\to0$ and $\xi\to0$ is
ambiguous, because the term
\beq
{-g_{ab}\over64\pi^2}m^2\left[m^2+\left(\xi-{1\over6}\right)R\right]\psi
\left({3\over2}-\lambda\right)\longrightarrow
{-g_{ab}\over1536\pi^2}{R^2\over 1+{\xi R\over m^2}}\label{ambi}
\eeq
gives different answers by approaching the
origin of the $(\xi,m^2)$ plane in different ways. It is intriguing that
in order to recover the result (\ref{caesar}), the limit
$m^2,\xi\to 0$ has to be taken
along a path such that
\beq
{\xi R\over m^2}\to-2. \label{path}
\eeq

The origin of the ambiguity can be traced back to the
contribution of the mode $L=0$ to $<T_{\mu\nu}>$ in the Euclidean vacuum. It
is easy to see that this contribution is given by
$$
{- R^2\over 1536\pi^2}{(\xi R+2m^2)\over (m^2+\xi R)}+O(m^2,\xi),
$$
and therefore it will vanish only if the limit is taken according to the path
(\ref{path}). This is equivalent to taking the limit $m^2,\xi\to 0$ in the
formal expression of the energy momentum tensor operator before taking the
vacuum expectation value.

\section{Conclusions and discussion}

We have used the Hadamard formalism to compute the renormalized
expectation value of the energy momentum tensor for a massless minimally
coupled field in de Sitter space in the two parameter family of O(4)
Hadamard vacua.
We find that this
tensor is not de Sitter invariant but only O(4) invariant (in disagreement
with the result of ref.\cite{alfo87},
which was subsequently corrected in ref.\cite{referee}).

We have also studied the de Sitter invariant state for the massless
minimally coupled field.
It is worth noting that such state is not a
Fock vacuum (indeed, Allen \cite{al85} has shown that for $m=\xi=0$ there
is no de Sitter invariant Fock vacuum): the discrete zero mode is not
quantized in terms of creation and anihilation operators, but rather
using the canonical position and momentum operators.
In particular, we have used it to derive a
covariant version of eq. (\ref{growth}). We find that the expectation value
of the square of the difference $\phi(x)-\phi(y)$ grows linearly with the
geodesic distance between $x$ and $y$,
for time-like separations which are large
compared with $H^{-1}$ [see eq.(\ref{result})].
The linear growth $D(x,y)\propto
H\tau$ has the same physical origin as the linear growth in time of eq.
(\ref{growth}) and it can be interpreted, along the same lines, as a
``Brownian motion'' of the field due to quantum fluctuations
(see e.g. ref.\cite{vi83}).

We have computed the renormalized expectation
value of the energy momentum tensor in the de Sitter invariant vacuum. We find
that the renormalized vacuum energy density ,$<T_{00}>_{ren}$,
is lower in this state than in any of the O(4) invariant
states. In this sense, only the de Sitter invariant state deserves to be
called vacuum.

The $O(4)$ invariant $<T_{\mu\nu}>_{ren}$, eq. (\ref{tmunu}), approaches
the de Sitter invariant value (\ref{caesar}) at
time-like infinity. Also, the dispersion $D(x,y)$ computed in
Section 4 using the de Sitter invariant state coincides with the limit
$\eta\to \pi$ of the dispersion computed in a $O(4)$ invariant state.
Therefore, the de Sitter invariant state can be seen as the
limit into which the $O(4)$ invariant states evolve at sufficiently late
times. This behaviour is familiar from the massive case, and it corresponds
to the fact that any excitations above the de Sitter invariant vacuum are
redshifted away by the exponential expansion.

After this paper was submitted, it was pointed out to us by the referee that
the result of Ref.\cite{alfo87} had already been corrected in
Ref.\cite{referee}.

\section*{Acknowledgements}

We would like to thank Larry Ford, Alex Vilenkin and Prof. V.F. M\"{u}ller
for useful conversations. The work of J.G. is supported by a Fulbright grant.

\appendix
\section*{Appendix A}
\setcounter{equation}{0}
\renewcommand{\theequation}{A\arabic{equation}}

For completeness, in this appendix we summarize the field quantization in the
Schr\"{o}dinger picture (see e.g. ref.\cite{ra85}).

In the Schr\"{o}dinger picture, $\hat\chi_{LM}$ and $\hat\pi_{LM}$ are time
independent operators satisfying the commutation relations
\beq
[\hat\chi_{LM},
\hat\pi_{L'M'}]=i\delta_{LL'}\delta_{MM'},\label{commutations}
\eeq
and acting on a Hilbert space of time dependent physical states $\Psi$.
In the `q' representation, such states are described by wave functionals
$\Psi(\{\chi_{LM}\},\eta)$ and the action of the operators is given by
$$
\hat\chi_{LM}\Psi=\chi_{LM}\Psi,
$$
$$
\hat\pi_{LM}\Psi=-i{\partial\over\partial\chi_{LM}}\Psi.
$$
The time evolution is governed by the Schr\"{o}dinger equation
\beq
\hat H\Psi=-i{\partial\over\partial\eta}\Psi,\label{schrodinger}
\eeq
where $\hat H$ is the Hamiltonian derived from the action (\ref{action2}),
with $\chi_{LM}$ and $\pi_{LM}$ replaced by its operator counterparts:
\beq
\hat H=\sum_{LM} {1\over2}\left[{\hat\pi_{LM}^2\over(H\sin\eta)^{-2}}+
(H\sin\eta)^{-2}\omega_L^2\chi_{LM}^2\right].
\label{hamiltonian}
\eeq
Note that throughout this appendix, $\chi_{LM}$ are not functions of $\eta$
(as they were in Section 2) but they are the time independent position
operators of the Schr\"{o}dinger picture (see e.g. \cite{re89}).

Factorizing the wave functional as
$$
\Psi=\prod_{LM}\Psi_{LM}(\chi_{LM},\eta),
$$
eq. (\ref{schrodinger}) separates into a set of Schr\"{o}dinger equations, one
for each individual mode
\beq
{1\over
2}\left[{-1\over(H\sin\eta)^{-2}}{\partial^2\over\partial\chi_{LM}^2} +
(H\sin\eta)^{-2}\omega_L^2\chi^2_{LM}\right]\Psi_{LM}=
-i{\partial\over\partial\eta}\Psi_{LM}.\label{ssss}
\eeq
These can be solved by using the ansatz
\beq
\Psi_{LM}=g_{LM}\exp{\left[{i\over 2}(H\sin\eta)^{-2}{\dot V_{LM}\over V_{LM}}
\chi_{LM}^2\right]},\label{ansatz}
\eeq
where $g_{LM}(\eta)$ and $V_{LM}(\eta)$ are unspecified functions.
Substituting (\ref{ansatz}) into eq.(\ref{ssss}) and collecting the terms
proportional to $\chi_{LM}^2$ one finds
\beq
\ddot V_{LM}-2\cot\eta\dot V_{LM}+\omega_L^2(\eta)V_{LM}=0,
\label{yup}
\eeq
so $V_{LM}$ must be a  solution of the field equation (\ref{eom}).
Collecting the terms which are independent of $\chi_{LM}$, one finds a
differential equation for $g_{LM}$ which can be solved immediately to yield
$$
g_{LM}=C_{LM}V^{-1/2}_{LM},
$$
where $C_{LM}$ is just an overall normalization constant. Choosing one
solution of (\ref{yup}) for each $L$ and $M$ specifies a particular quantum
state. In order to know what set of solutions $\{V_{LM}\}$ corresponds to
the de Sitter invariant quantum state defined in Section 2, one has to
impose that the wave functional be anihilated by the operators $a_{LM}$
associated with the set of modes that defines such vacuum, eq. (\ref{modes}):
$$
a_{LM}\Psi=\left[U^*_{LM}{\partial\over\partial\chi_{LM}}
-i{\dot U^*_{LM}\over(H\sin\eta)^2}\chi_{LM}\right]\Psi=0,
$$
where we have inverted (\ref{customary}) to express $a_{LM}$ in terms of
$\hat\chi_{LM}$ and $\hat\pi_{LM}$. Clearly, this conditions are satisfied
if and only if
$$
V_{LM}=U^*_{LM}.
$$

In summary, the de Sitter invariant wave functional is given by
\beq
\Psi=\prod_{LM}(2\pi)^{-1/4}U_{LM}^{-1/2}\exp{\left[{i\over 2}(H\sin\eta)^{-2}
{\dot U^*_{LM}\over U^*_{LM}}\chi^2_{LM}\right]},\label{wavefunctional}
\eeq
with $U_{LM}$ given by (\ref{modes}). It can be checked that this
wave-functional is anihilated by the operator generators of the de Sitter
group \cite{re89}, and is thus de Sitter invariant.
Note also that this wave functional
is properly normalized, in the sense that
$$
\int^{\infty}_{\infty}\prod_{LM}d\chi_{LM}|\Psi(\{\chi_{LM}\},\eta)|^2=1.
$$

Note that the case $m^2=\xi=0$ is special. From (\ref{modes}) we find that
$U_0$ becomes constant in the massless minimally coupled limit,
$$
U_0=A_0\left(-\sqrt{2\over\pi}\right),
$$
so the Wronskian condition can not be satisfied and the normalization
constant $A_0$ becomes infinite [see eq.(\ref{coefficients})].
Such infinity can be understood by noticing that,
since
$$
\lim_{m^2,\xi\to 0}{\dot U_0\over U_0}= 0,
$$
the wave functional becomes
independent of $\chi_0$
$$
\pi_0\Psi=-i{\partial\over\partial\chi_0}\Psi=0,\quad(m^2=\xi=0),
$$
and therefore
$\Psi$ is not normalizable in the discrete sense (which is natural
for an eigenstate of momentum).

To conclude, let us study in more detail the $L=0$ term of the
Schr\"{o}dinger equation (\ref{schrodinger}). Using the notation $\psi\equiv
\Psi_{L=0}$ we have
$$
-{1\over 2}{\partial\over\partial\chi_0^2}\psi={-i\over H^2}{\partial\over
\partial \tilde t}\psi,
$$
where we have introduced the new time variable
$$\tilde t\equiv {1\over 2}(\eta-{1\over 2}\sin 2\eta-{\pi\over 2}).$$
In this notation the basic solutions are the eigenstates of momentum that
we discussed in Section 4,
$$
\psi_p(\chi_0)\propto e^{i(pq-p^2\tilde t)}
$$
with $q\equiv \sqrt{2}H^{-1}\chi_0$ [see Eq. (\ref{jacobi})]. For $\tilde
t=0$ these are the Heisenberg wave functions (\ref{pws}).

The wave packet (\ref{packet}) is just a superposition of these modes, and
its time evolution can be found in any  elementary textbook. It represents
a gaussian wave-packet that spreads in time. Noting that
$<\alpha|P^2|\alpha>=2\alpha^2$ and $<\alpha|Q^2|\alpha>=(2\alpha^2)^{-1}$
we have, from (\ref{jacobi}),
$$
<\chi_0^2>_{\alpha}=H^2\left[{1\over 4\alpha^2}+4\alpha^2\tilde t^2\right].
$$
Since the range of $\tilde t$ is finite, $\tilde t\in[-\pi/4,\pi/4]$, the
expectation value of $\chi_0^2$ does not grow unbounded, but reaches a
constant in the asymptotic past and future. Therefore the
asymptotic growth in time
of $<\phi^2>$ in de Sitter space is due to the $L>0$ modes.

This behaviour is
somewhat different from that of the theory of a massless field
on a compact toroidal flat spacetime which we briefly discussed in Section
4. There, the contribution of the $L=0$ mode to $<\phi^2>$ also has
a term proportional to $<p_0^2>t^2$.
However, in that case, $t$ is the Minkowski time.
If we choose a state with $<p_0^2>\neq 0$ then $<\phi^2>\propto t^2$ grows
unbounded as time increases due to the $L=0$ contribution alone.
On the other hand, for the ground state
$<p_0^2>=0$ but $<x_0^2>=\infty$ and therefore $<\phi^2>$ is infinite,
just like in the de Sitter invariant state studied in this paper.


\begin{thebibliography}{999}





\bibitem{bida82} See, e.g. N.D. Birrell and P.C.W. Davies
{\em ``Quantum Fields in Curved Space''}, Cambridge University Press,
Cambridge (1982).

\bibitem{cheta68} N.A. Chernikov and E.A. Tagirov,
Ann. Inst. H. Poincare, {\bf IX}, 109 (1968); E.A. Tagirov,
Ann. Phys. {\bf 76}, 561 (1973).

\bibitem{buda78} T.S. Bunch and P.C.W. Davies, Proc. R. Soc. London {\bf A360},
117 (1978).

\bibitem{blgu} For a review on inflation, see e.g.
S.K. Blau and A.H. Guth, in {\em 300 Years of Gravitation},
edited by S. Hawking and W. Israel, Cambridge University Press, Cambridge
(1987) .

\bibitem{vifo82} A. Vilenkin and L.H. Ford, Phys. Rev. {\bf D26}, 1231
(1982).

\bibitem{al85} B. Allen, Phys. Rev. {\bf D32}, 3136 (1985).

\bibitem{alfo87} B. Allen and A. Folacci, Phys. Rev. {\bf D35}, 3771
(1987).

\bibitem{po91} D. Polarski, Phys. Rev. {\bf D43}, 1892 (1991).

\bibitem{ra85} B. Ratra, Phys. Rev. {\bf D31}, 1931 (1985).

\bibitem{jackiw} R. Floreani, C.T. Hill and R. Jackiw , Ann. Phys. {\bf
175}, 345 (1987).

\bibitem{scsp76} C. Schomblond and P. Spindel, Ann. Inst. Henri Poincar\'e
{\bf A25}, 67 (1976).

\bibitem{re89} I. Redmount, Phys. Rev. {\bf D40}, 3343 (1989).

\bibitem{befo86} D. Bernard and A. Folacci, Phys. Rev. {\bf D34}, 2286 (1986).

\bibitem{fr75} F.G. Friedlander, {em The wave equation on a curved spacetime},
Cambridge University Press, Cambridge (1975).

\bibitem{dewitt63} B.S. De Witt, in {\em Les Houches 1963}, edited by
C. De Witt and B.S. De Witt (Gordon and Breach, New York, 1963).

\bibitem{brot86} M.R. Brown and A.C. Ottewill, Phys. Rev. {\bf D34}, 1776
(1986).


\bibitem{dewitt83} B.S. De Witt, in {\em Les Houches 1983}, edited by
C. De Witt and B.S. De Witt (Gordon and Breach, New York, 1983).

\bibitem{fopa89} L.H. Ford and C. Pathinayake, Phys. Rev. {\bf D39},
3642 (1989).

\bibitem{gavi91b} J. Garriga and A. Vilenkin, ``Quantum fluctuations on domain
walls, strings and vacuum bubbles'' (submitted to Phys. Rev. D.)

\bibitem{rajaraman} R. Rajaraman, {\em ``Solitons and Instantons''},
North-Holland, Amsterdam (1982).


\bibitem{collins} Collins, {\em ``Renormalization''}, Cambridge University
Press, Cambridge (1984).

\bibitem{vi83} A. Vilenkin, Nucl. Phys. {\bf B226}, 527 (1983).

\bibitem{ha23} J. Hadamard, {\em ``Lectures on Cauchy's Problem in linear
partial differential equations''} (Yale University Press, New Haven, CT, 1923).

\bibitem{referee} A. Folacci, J. Math. Phys. {\bf 32}, 2828 (1991);E{\bf
33}, 1932 (1992).


\end{thebibliography}
\end{document}